\documentclass[fleqn,10pt]{wlscirep}
\usepackage[utf8]{inputenc}
\usepackage[T1]{fontenc}

\usepackage{amsmath,amsthm,amssymb,amsfonts}
\usepackage{comment}
\usepackage{graphicx}
\usepackage{pdfpages}
\usepackage{algorithm}
\usepackage{algpseudocode}
\usepackage{appendix}
\usepackage{siunitx}  
\usepackage{booktabs} 
\usepackage{tabularx} 
\usepackage{bbm}

\newcommand{\pder}[2][]{\frac{\partial#1}{\partial#2}}

\usepackage[utf8]{inputenc}
\usepackage[english]{babel}

\title{Active Disruption Avoidance and Trajectory Design for Tokamak Ramp-downs with Neural Differential Equations and Reinforcement Learning}

\author[1,2,*]{Allen M. Wang}
\author[2]{Oswin So}
\author[2]{Charles Dawson}
\author[1,3]{Darren Garnier}
\author[1]{Cristina Rea}
\author[2]{Chuchu Fan}
\affil[1]{Plasma Science and Fusion Center, Massachusetts Institute of
    Technology, Cambridge, MA 02410, USA}
\affil[2]{Laboratory for Information and Decision Systems, Massachusetts
    Institute of Technology, Cambridge, MA 02410, USA}
\affil[3]{OpenStar Technologies LLC., Wellington, Wellington 6035, New Zealand}

\affil[*]{Corresponding Author: allenw@mit.edu}

\begin{abstract}
The tokamak offers a promising path to fusion energy, but plasma disruptions pose a major economic risk, motivating considerable advances in disruption avoidance. This work develops a reinforcement learning approach to this problem by training a policy to safely ramp-down the plasma current while avoiding limits on a number of quantities correlated with disruptions. The policy training environment is a hybrid physics and machine learning model trained on simulations of the SPARC primary reference discharge (PRD) ramp-down, an upcoming burning plasma scenario which we use as a testbed. To address physics uncertainty and model inaccuracies, the simulation environment is massively parallelized on GPU with randomized physics parameters during policy training. The trained policy is then successfully transferred to a higher fidelity simulator where it successfully ramps down the plasma while avoiding user-specified disruptive limits. We also address the crucial issue of safety criticality by demonstrating that a constraint-conditioned policy can be used as a trajectory design assistant to design a library of feed-forward trajectories to handle different physics conditions and user settings. As a library of trajectories is more interpretable and verifiable offline,  we argue such an approach is a promising path for leveraging the capabilities of reinforcement learning in the safety-critical context of burning plasma tokamaks. Finally, we demonstrate how the training environment can be a useful platform for other feed-forward optimization approaches by using an evolutionary algorithm to perform optimization of feed-forward trajectories that are robust to physics uncertainty.
\end{abstract}
\begin{document}

\flushbottom
\maketitle
%
%
\thispagestyle{empty}

\section*{Introduction}

Plasma disruptions, events where control of the plasma is lost, stand as a
major barrier to realizing economical fusion energy through tokamak devices.
Unmitigated disruptions in existing tokamak pilot plant designs, such as ARC, may disrupt machine operations
for months or longer \cite{maris2023impact}. Even at the energies of upcoming
burning plasma devices such as SPARC and ITER, disruptions represent a major
challenge, driving costly design requirements and posing risks of major delays to tokamak operations \cite{de2016requirements, sweeney2020mhd}. To
overcome the challenge of disruptions, considerable advances are needed in avoidance, mitigation, resilience, and recovery (AMRR) techniques. Given that
``mitigation'' is crucial for avoiding the most expensive consequences of
disruptions, a considerable body of research has evolved developing both
mitigation actuators, such as massive gas injection
(MGI)\cite{whyte2002mitigation}, and disruption prediction algorithms to
trigger said mitigation actuators \cite{rea2019real,montes2019machine,pau2019machine}. While robust, reliable mitigation is crucial for investment protection, it is still expected to induce considerable stress on tokamak components and should
be considered a method of last resort \cite{eidietis2018implementing}. Thus,
\textit{disruption avoidance} would be preferable. 

In this paper, we define \textit{active disruption avoidance}  as any real-time technique where the
plasma control system (PCS) intelligently commands actuators to drive the
plasma state away from instability limits and either returns to nominal
operations or de-energizes the plasma to a sufficiently low energy such that a
disruption is inconsequential. We focus specifically on the ``ramp-down'' phase, where the
plasma is de-energized. The primary figure of merit for a successful ramp-down is often the plasma current $I_p$
at which the plasma terminates as both structural loading and the risk posed by runaway electrons scale with $I_p$ \cite{hender2007mhd}. In recent years, works have developed both
Sequential Quadratic Programming (SQP) and Bayesian Optimization approaches for
designing fast feed-forward ramp-down trajectories that are programmed into the
machine \cite{teplukhina2017simulation,van2023scenarioA,van2023scenarioB}. However, many challenges remain unaddressed such as demonstrating scalability to allow for the generation of many trajectories to handle different scenarios, and the optimization of trajectories that are \textit{robust} to physics uncertainty. In addition, \textit{active disruption avoidance}, remains an
under-developed area of research with calls for further advancement
\cite{boyer2021toward,boozer2021plasma}.

The problem setting of ramp-downs and active disruption avoidance faces the considerable challenge of physics uncertainty. While recent work successfully demonstrated the
application of reinforcement learning (RL) to the problem of shape control\cite{degrave2022magnetic}, the physics relevant to the
problem, ideal Magnetohydrodynamics (MHD), is arguably the most well-understood
and well-simulated aspect of tokamak plasma physics.
Disruption avoidance, however, is  more challenging as it involves complex
physical phenomena such as turbulent transport for which accurate, full-shot
simulation from physics-based principles remains an open research problem. The
highest fidelity simulations of core turbulence performed today with nonlinear
gyrokinetics take millions of CPU-hours to simulate a steady-state condition
and, even then, require boundary condition assumptions on quantities such as
the pedestal pressure and impurity densities
\cite{rodriguez2022nonlinear}. The extreme computational requirements of
physics-based simulations stands in stark contrast to tokamak operations
requirements. When unexpected physics or technical issues arise during tokamak
operations, dynamics models, trajectories, and controllers need to be updated
as soon as possible to ensure subsequent discharges are successful. As a
consequence, scenario and control design is often performed with highly
simplified dynamics models through empirical power-law scalings and linear
systems obtained with techniques from classical system identification
\cite{walker2020introduction}. In fact, both the ITER and SPARC nominal
operating scenarios were designed using empirical plasma operational contour
(POPCON) scalings with scenario validation performed with higher fidelity
simulations
\cite{creely2020overview,rodriguez2020predictions,international2001iaea}.

This challenge has motivated the development of data-driven dynamics models of
existing devices for the purposes of reinforcement learning, but standard sequence-to-sequence models in the deep learning toolbox, such as recurrent neural networks
(RNNs) and transformers, are generally not sample efficient. This is a major challenge as burning plasma tokamaks must work reliably with
as few shots of training data as possible, given the high cost of disruptions; existing works often use more than $10^4$ shots of data
\cite{abbate2021data,char2023offline}. Sample-efficient dynamics modelling methods are needed.

\textbf{Statement of Contributions}: This paper aims to advance the state-of-the-art on three fronts: sample-efficient dynamics model learning, active disruption avoidance, and optimizing feed-forward trajectories with robustness to physics uncertainty.  To address the model learning problem, we train a simple hybrid dynamics model that contains both physics and machine learning components on a relatively small dataset of simulations of SPARC PRD ramp-downs with the control-oriented simulator RAPTOR \cite{felici2011thesis,felici2011real}. This dynamics model is fully implemented in the machine learning framework JAX\cite{jax2018github}, and is thus fully differentiable and GPU-capable, enabling massive parallelization across physics assumptions. We then wrap this dynamics model in an OpenAI Gym\cite{brockman2016openai} environment to create what we call ``PopDownGym''. We use PopDownGym to address the matters of feed-forward trajectory optimization and active disruption avoidance in a unified reinforcement learning (RL) approach. As a first step towards experimental demonstration of active disruption avoidance, we train a policy on PopDownGym with Proximal Policy Optimization (PPO) and demonstrate its transfer to RAPTOR where it successfully ramps down a simulated plasma while avoiding user specified limits. We address the issue of safety criticality by demonstrating how a constraint-conditioned policy can be used to design a library of trajectories to handle different conditions by performing policy rollout on simulators under different physics settings. Finally, we bridge the gap between reinforcement learning and more conventional trajectory optimization approaches by using the RL training environment to perform robust optimization of a feed-forward trajectory with an evolutionary algorithm. 
\section*{Background}
\subsection*{Reinforcement Learning and Optimal Control}
This section aims to provide a brief overview of reinforcement learning and show how it can be applied to the problem of trajectory optimization, which is typically approached from the perspective of optimal control. Reinforcement learning is typically concerned with solving Partially Observable
Markov Decision Processes (POMDPs). A POMDP is defined by states $\mathbf{x}$
in a state space $\mathcal{X}$, actions $\mathbf{a}$ in an action space
$\mathcal{A}$, a reward function mapping states and actions to a real-valued
scalar $r(\mathbf{x}, \mathbf{a}):
    \mathcal{X}\times\mathcal{A}\rightarrow\mathbb{R}$ and a, possibly
stochastic,
dynamics model that maps state and action at time $t$ to states at the next time step: $
    \mathbf{x}_{t+1} = f(\mathbf{x}_t,\mathbf{a}_t)$. In a POMDP, the state is not observable to the
control policy, it instead receives an observation vector from an observation function $O$ that maps the
underlying state vector to the observation space,
$\mathcal{O}: \mathbf{o}_{t} = O(\mathbf{x}_t)$. The full problem statement of a POMDP is thus given by:
\begin{subequations}
    \begin{align}
        \max_{\mathbf{a}_{0:T}} & \sum_{t=0}^T \gamma^t\mathbb{E}[r(\mathbf{x}_t,\mathbf{a}_t)]\label{eq:pomdp_objective}
        \\
        s.t. \quad              & \mathbf{x}_{t+1} =
        f(\mathbf{x}_t,\mathbf{a}_t)\quad \forall t
        \\
                                & \mathbf{o}_{t} = O(\mathbf{x}_t)\quad\forall
        t
        \\
                                & \mathbf{x}_t\in\mathcal{X}\quad\forall t
        \\
                                & \mathbf{a}_t\in\mathcal{A}\quad\forall t
    \end{align}
\end{subequations}
In a POMDP, the reward function is usually engineered to achieve the desired
outcome. For example, in the context of video games, the reward function is
designed to express wins and losses. In the context of the ramp-down problem
addressed by this paper, we design the reward function to reward de-energizing the plasma while avoiding disruptive and machine limits. The
discount factor $\gamma\in[0,1]$ often augments the reward function to reduce
the weight of future rewards and prioritize present rewards. Solutions to
POMDPs typically take one of two forms: either a control policy or a
trajectory. A \textit{control policy} is a function mapping observations to actions
$\mathbf{a}_t = \pi(\mathbf{o}_t)$.
In the modern deep reinforcement learning paradigm, it is usually implemented as a deep neural network.

Another possible solution type is an open-loop \textit{trajectory}. For a given initial
state, $\mathbf{x}_0$, and corresponding observation, $\mathbf{o}_0$, an
\textit{action trajectory} is a sequence of actions $[\mathbf{a}_0,
            \mathbf{a}_1, ..., \mathbf{a}_{T}]$ over time. Observe that given
access to the
dynamics function $f$ and observation function $O$, one can use the initial
state and an action trajectory to generate a \textit{state trajectory} by
recursive application of the dynamics function:
\begin{subequations}
    \begin{align}
        \mathbf{x}_{1}   & = f(\mathbf{x}_0, \mathbf{a}_0) \\
        \mathbf{x}_{2}   & = f(\mathbf{x}_1, \mathbf{a}_1) \\
                         & \vdots\notag                    \\
        \mathbf{x}_{T+1} & = f(\mathbf{x}_T, \mathbf{a}_T)
    \end{align}
\end{subequations}
and then an \textit{observation trajectory} can be generated by applying $O$ to
the state trajectory. The computational problem of finding optimal trajectories
is often referred to as \textit{trajectory optimization} and has largely been
advanced by the field of \textit{optimal control}. In fact, the POMDP formulation is directly analogous to standard formulations of \textit{stochastic optimal control} \cite{bertsekas2019reinforcement}.

A control policy can be viewed as a more general solution than a trajectory as,
given some initial state, the control policy can be used along with the
transition function to generate a trajectory through a process known as
\textit{policy rollout}. The idea is to recursively map the initial state to an
observation, feed the observation into the policy to determine an action, and
then apply the state and action to the simulator $f$:
\begin{subequations}
    \begin{align}
        \mathbf{a}_0 & = \pi(O(\mathbf{x}_0)) \quad \mathbf{x}_1 =
        f(\mathbf{x}_0, \mathbf{a}_0)                                  \\
        \mathbf{a}_1 & = \pi(O(\mathbf{x}_1)) \quad \mathbf{x}_2 =
        f(\mathbf{x}_1, \mathbf{a}_1)                                  \\
                     & \vdots\notag
        \\
        \mathbf{a}_T & = \pi(O(\mathbf{x}_T)) \quad \mathbf{x}_{T+1} =
        f(\mathbf{x}_T, \mathbf{a}_T)
    \end{align}
\end{subequations}
While control policies trained through RL are often thought of as products
for real-time control, the process of rollout can also be used to generate
trajectories offline through simulation. The other key distinction
between the RL and optimal control frameworks is that the latter explicitly supports
constraints. However, as shown in the Methods section, constraints can be
implemented in reward functions via penalty terms. In fact, many nonlinear
optimization methods used in optimal control implement constraints as cost
terms under the hood with methods such as augmented lagrangians and barrier
functions \cite{nesterov2018lectures,bertsekas2014constrained}.

\subsection*{Challenges of Plasma Dynamics Modelling}
In the context of plasma control, defining the dynamics function $f$ and
observation function $O$ are considerable challenges. While accurate full-shot
simulation of plasmas from first principles is itself an open research problem,
application of RL techniques generally requires stable, massively parallelized
simulation with randomized physics and thus imposes even more stringent
software requirements. These challenging requirements have motivated work on
developing neural network dynamics models trained on machine data\cite{abbate2021data,char2023offline}. However, the
relative paucity of available fusion data makes sample-efficiency and
robustness of these models a considerable challenge.

At the same time, reactor and control design are often  performed
using relatively simple 0-D models which are used to find design points that
are then validated using higher fidelity transport simulations. In the context
of reactor design, such models are often referred to as Plasma Operating
Contour (POPCON) models and include well-established physics, such as
reactivity curves, with empirical scalings, such as $\tau_E$ scalings \cite{creely2020overview,frank2022radiative}. Despite
their simplicity, such models have produced predictions of key performance
metrics such as energy gain and $\tau_E$ that are in line with nonlinear
gyrokinetic simulations \cite{}. In the context of control design, classical
\textit{system identification} techniques are often employed to fit linear
dynamical systems to empirical data; controllers are then designed on these
linear models \cite{wang2018system}. Despite the complexity of plasma dynamics,
such techniques have proven effective for control design in certain cases
\cite{felici2011real}.

\subsection*{Neural Differential Equations and Differentiable Simulation}
While classical system identification techniques have yielded successes,
further advances are needed. Fortunately, recent advances made by the
scientific machine learning community in differentiable simulation and
neural differential equations offers a powerful new toolbox for complex
nonlinear system identification \cite{kidger2022neural,chen2018neural,rackauckas2020universal}. This work applies such advances to train
POPCON-like dynamics models that combine physics structure with machine
learning to predict dynamics from relatively few samples. In contrast to
standard POPCON models, the model is time-dependent and its free parameters can
be updated to new empirical data, and, in contrast to classical system
identification techniques, the model can be highly nonlinear.

Consider the following nonlinear dynamical system where $f_\theta$ is a neural
network with free parameters $\theta$:
\begin{align}
    \frac{d\mathbf{x}}{dt} = f_\theta(\mathbf{x})
\end{align}
Such an equation is known as a neural differential equation (NDE). Both training and inference of
NDEs involves numerically integrating the neural network from its
initial condition $\mathbf{x}(0)$ across some time horizon to some terminal
value. While standard feed-forward neural networks have the following input-output
relationship:
\begin{align}
    \mathbf{y} = \operatorname{NN}_\theta(\mathbf{x})
\end{align}
NDEs have the following input-output relationship:
\begin{align}
    \mathbf{y} = \operatorname{diffeqsolve}(f_\theta, \mathbf{x}_0)
\end{align}
where $\operatorname{diffeqsolve}$ is a numerical integration scheme such as explicit or
implicit Runge-Kutta methods. Then, given data $\hat{\mathbf{y}}$, and a loss
function defining error between simulation and data
$\mathcal{L}(\mathbf{y},\hat{\mathbf{y}})$, \textit{adjoint methods} \cite{chen2018neural,kidger2022neural} can be
used to efficiently compute the gradient of this loss function with respect to
the model parameters $\nabla_\theta\mathcal{L} (\mathbf{y}, \hat{\mathbf{y}})$. While we have thus centered this discussion around the case where $f_\theta$ is
a neural network, the same methodology applies to all functions $f_\theta$ that
are differentiable with respect to $\theta$. Historically, the differentiability requirements have limited the complexity of $f_\theta$. However, modern automatic differentiation frameworks allow for highly complex choices of $f_\theta$ that can also include physics. This more general setting is often referred to
as \textit{differentiable simulation} and enables the development of $f_\theta$
models that contain both physics and neural network components.

\section*{Results}\label{sec:results}
We use the simulator RAPTOR configured to the SPARC primary reference discharge (PRD)\cite{creely2020overview}, an 8.7 MegaAmpere (MA) H-mode scenario projected to reach $Q\approx 11$, as our testbed. We first generate a dataset of 481 ramp-down simulations with varying plasma current and auxiliary heating ramp-rates. 336 of these ramp-down simulations are then used to train the PopDownGym dynamics model while the rest are used for validation and test.

As shown in Table \ref{tab:obs_action_ranges}, the policy receives an observation from an eight dimensional observation space to select an action vector from a four dimensional action space. It has the goal of ramping down the plasma current to below two MA, a level existing devices have safely operated at \cite{hender2007mhd}, while avoiding user-specified constraints on eight quantities which the device operator will want to limit. As discussed further in the Methods section, we train the control policy to accept user-specified constraint limits as inputs. This allows the constraints to be adjusted at inference time \textit{without retraining}.

This section highlights two uses cases for the trained policy: 1) as a trajectory design assistant to design feed-forward trajectories for a wide range of physics conditions, and 2) as a real-time supervisory controller for actively avoiding disruptive limits. As a step towards demonstrating 2) experimentally, we transfer the control policy to RAPTOR. Finally, we demonstrate that the same gym environment can be leveraged to perform \textit{robust optimization} of trajectories (i.e. optimization of feedforward trajectories that will succeed under a range of physics conditions).
\begin{table}[!h]
\centering
\renewcommand{\arraystretch}{1} 
\begin{tabularx}{\textwidth}{@{}cXcc@{}} 
\toprule
\textbf{Parameter} & \textbf{Description} & \textbf{Min} & \textbf{Max} \\
\midrule
\multicolumn{4}{c}{\textbf{Observation Space}} \\
\addlinespace
$l_i$ & Normalized internal inductance & 0.5 & 6.0 \\
$I_p$ & Plasma current [MA] & 1.0 & 9.0 \\
$V_R$ & Voltage (Eq. 42 in \cite{romero2010plasma}) [V] & -5.0 & 5.0 \\
$W_{th}$ & Plasma energy [J] & $10^5$ & $3\times 10^7$ \\
$n_i$ & Fuel density \([10^{19} \text{m}^{-3}]\) & 1.0 & 30.0 \\
$P_{aux}$ & Aux. heating power [MW] & 0.0 & 25.0 \\
$g_s$ & Geometry parameter & 0.0 & 1.0 \\
\midrule
\multicolumn{4}{c}{\textbf{Action Space}} \\
\addlinespace
$dI_p/dt$ & Change in plasma current [MA/s] & -3.0 & -0.5 \\
$dP_{aux}/dt$ & Change in aux. heating [MW/s] & -5.0 & 5.0 \\
$u_{fuel,19}$ & Fueling rate [$10^{19}$/s] & 0.0 & 10.0 \\
$dg_s/dt$ & Geometry change rate [1/s] & 0.0 & 1.0 \\
\midrule
\multicolumn{4}{c}{\textbf{Constrained Variables}} \\
\addlinespace
$l_i$ & Normalized inductance & 2 & 3 \\
$n_{g,frac}$ & Greenwald fraction & 0.5 & 0.8 \\
$\beta_N$ & Normalized beta & 0.015 & 0.028 \\
$\beta_p$ & Poloidal beta & 0.25 & 0.4 \\
$dB_v/dt$ & Vertical field change rate [T/s] & 0.2 & 0.4 \\
$dW_{th}/dt$ & Energy change rate [W] & $2\times 10^7$ & $7\times 10^7$ \\
$\Gamma$ & Shafranov coefficient & 3.4 & 3.6 \\
$\iota_{95}$ & Inverse safety factor & 0.35 & 0.45 \\
\bottomrule
\end{tabularx}
\caption{Observation and action spaces with bounds. Variables out of bounds terminate the episode. Constrained values range for policy training.}
\label{tab:obs_action_ranges}
\end{table}
\subsection*{Policies as Trajectory Design Assistants}
\begin{figure}[!h]
    \centering
    \includegraphics[width=\linewidth]{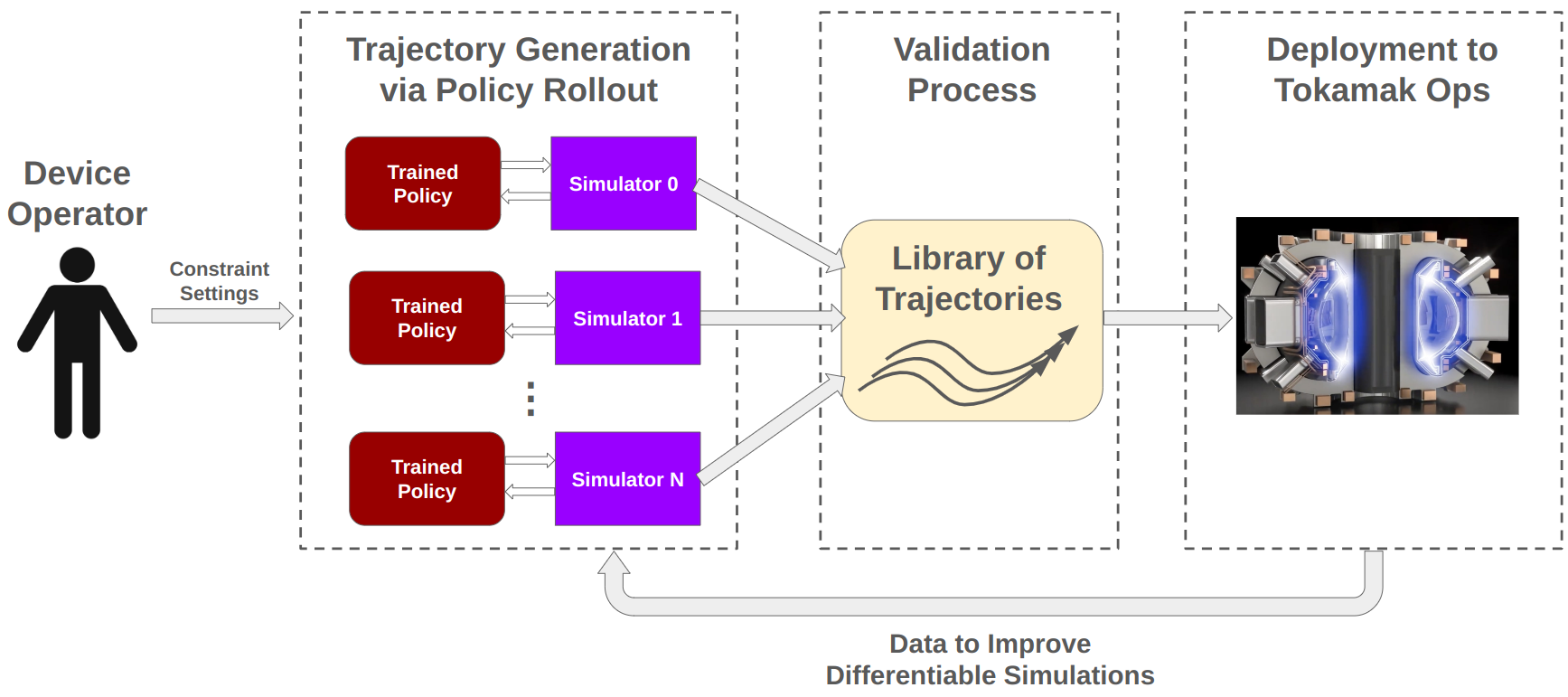}
    \caption{Simplified diagram of the proposed workflow for using a trained control policy as a trajectory design assistant.}
    \label{fig:assistant_workflow}
\end{figure}

It is desirable for burning plasma devices such as SPARC and ITER to achieve their
core mission with as few shots and disruptions as possible. In this context,
the
application of a learned control policy, which generally takes a non-trivial
amount of trial and error to transfer to reality, is not desirable without extensive validation and operational history on existing devices.
However,
we demonstrate that the learned control policy can still be used as a
trajectory
design assistant to design a \textit{library} of feedforward trajectories to handle a variety of different scenarios. These feedforward trajectories can
then be validated against
higher fidelity simulations before being programmed into the machine. Figure \ref{fig:assistant_workflow} depicts this workflow.

\subsubsection*{Constraint-Conditioned Policies}
Due to uncertainty about disruptive limits, a key feature in a useful
trajectory
design assistant is the ability for the user to adjust the constraints at
inference time. For example, there is uncertainty about what values of
$\beta_p$ may be correlated with dangerous neoclassical tearing modes (NTMs)
and what values of $l_i$ or the Shafranov coefficient may be correlated with
a vertical displacement event (VDE). After new information about such limits
is acquired through tokamak operations, it is desirable to redesign trajectories given this new
information
as quickly as possible to enable operational continuity.

\begin{figure}[!h]
    \centering
    \includegraphics[width=\linewidth]{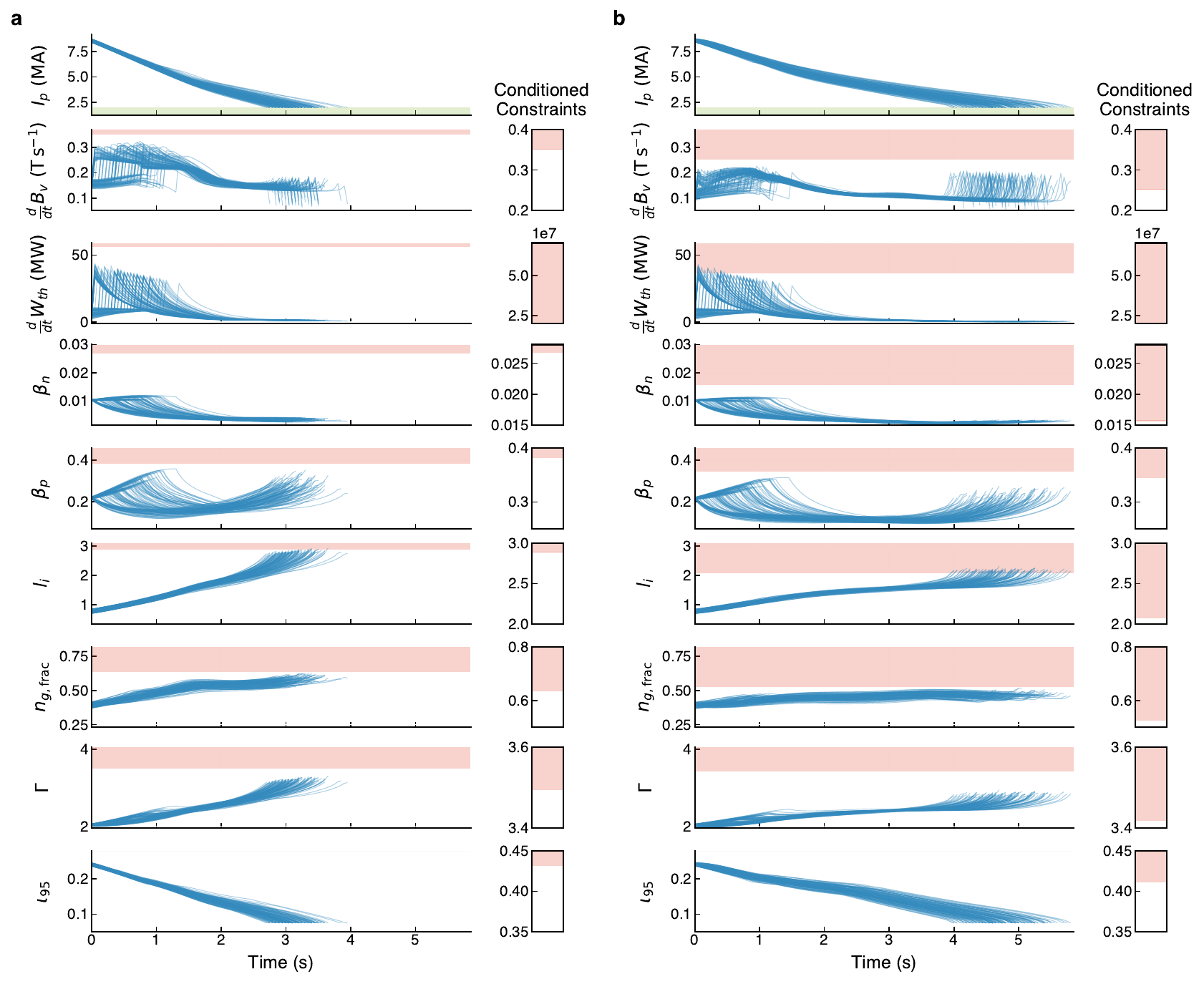}
    \caption{\textbf{Constraint-conditioned library of trajectories. } Library of trajectories generated under flexible (\textbf{a}) and more stringent (\textbf{b}) constraints. The green shaded region in the $I_p$ plots denotes the target plasma current of below \qty{2}{MA}, while the red shaded regions denote the constraint values. The constraint values conditioned on for each scenario in relation to the minimum and maximum values used for training are visualized as red bars on the right of each plot.}
    \label{fig:library}
\end{figure}

To address this challenge, we train a \textit{constraint-conditioned} policy
that allows the user to specify the constraint boundaries at inference time.
Figure \ref{fig:library}
shows a library of trajectories generated with two sets of constraint settings; one more relaxed, the other more stringent. The new
trajectories are generated without any policy re-training, thus, they can be rapidly generated with a single policy rollout. In our demonstration, this is done within seconds. In Figure \ref{fig:library}, the policy mostly does a good job of trying to satisfy the constraints, with a small amount of violation in $l_i$ at the end of the trajectory in the more stringent constraint case. We find that the policy struggles to satisfy the more stringent $\frac{dW_{th}}{dt}$ requirement, even in experiments where we tried increasing the penalty of constraint violation. We hypothesize this is due to it being physically infeasible to avoid the large stored energy change caused by the confinement regimes transition in certain cases.

\subsection*{Active Disruption Avoidance: Sim2Sim Policy Transfer to RAPTOR}
\begin{figure}
    \centering
    \includegraphics[width=0.9\linewidth]{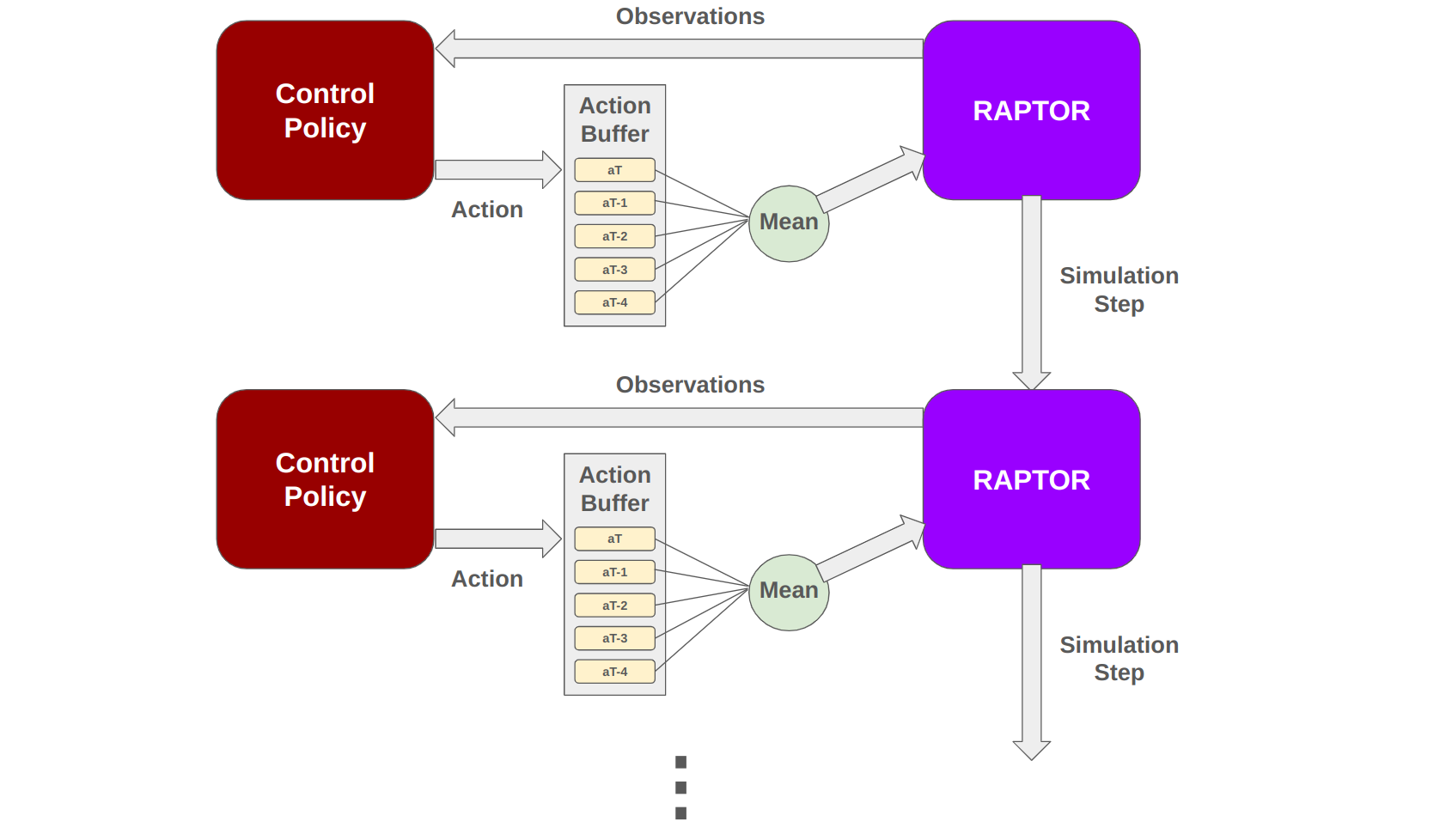}
    \caption{Diagram depicting the sim2sim transfer setup.}
    \label{fig:sim2sim_diagram}
\end{figure}

One use case for this learned policy is to serve as a supervisory
controller that uses real-time observations of the plasma state to make
decisions about how to ramp-down the plasma. As a  step towards experimentally demonstrating
such a capability, we demonstrate the successful transfer of the control policy to RAPTOR as a supervisory controller. Figure \ref{fig:sim2sim_diagram} shows the setup for our sim2sim demonstrations. The control policy receives observations from RAPTOR at each time and decides on an action. To smooth out the effects of numerical spikes from the simulation, we employ an action buffer such that the average of the five latest actions is input into RAPTOR. Similar techniques are employed in other reinforcement learning applications where noise and outlier measurements need to be handled \cite{hwangbo2019learning}.

\begin{figure}[!h]
    \centering
    \includegraphics[width=\textwidth]{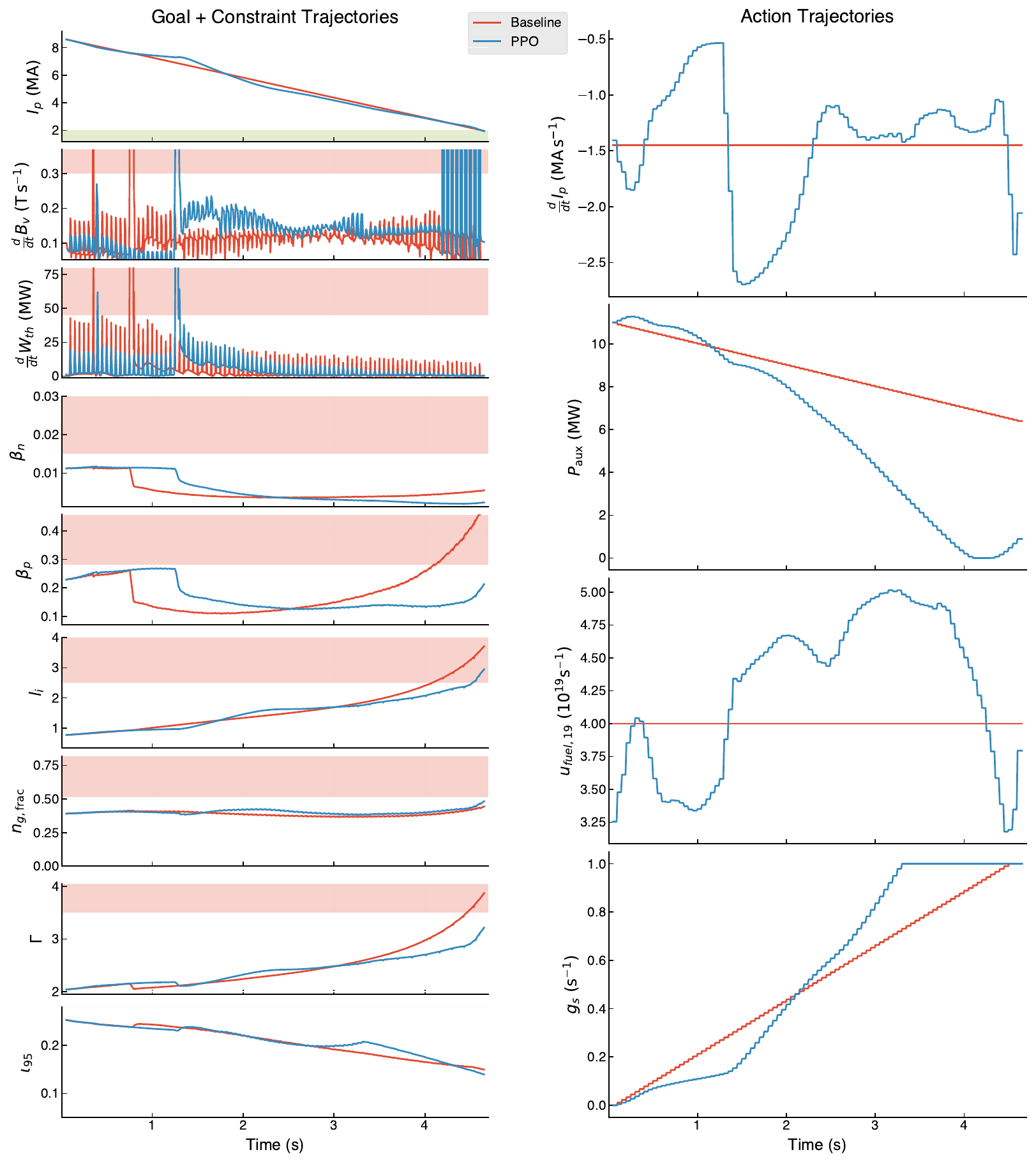}
    \caption{ Comparison of RAPTOR simulation results from a naive baseline feed-forward trajectory (red) against the PPO trained policy running in closed loop (blue). The PPO policy clearly yields a considerable reduction in constraint violation. While the nominal SPARC PRD ramp-down has a constant ramp-rate of 1MA/s, our baseline was selected to have the same average ramp-rate as the PPO policy to provide a better comparison between the two cases. Note that while the nominal action space is the rates of change of $P_{aux}$ and $g_s$, the plot shows the time-integrated values for interpertability.}
    \label{fig:sim2sim_comparison}
\end{figure}

Figure \ref{fig:sim2sim_comparison} shows results of one such experiment compared against a naive baseline feedforward trajectory, where many of the constraints are violated. The policy overall does a good job of avoiding the constraint boundaries, but there are imperfections. Like in the stringent constraint case in Figure \ref{fig:library}, we observe a small amount of $l_i$ constraint violation at the end of the ramp-down. In the action space, this can be attributed to the policy deciding to extremely quickly ramp the plasma current once it is close to its goal, which is not necessarily undesirable. We also observe spikes in $\frac{d}{dt}W_{th}$ and $\frac{d}{dt}B_v$ at the H to L mode transition, which we attribute to the numerics of boundary condition switching at the H to L mode transition. After the spike, the values quickly settle to values within the specified constraints. Similar spikes in $\frac{d}{dt}B_v$ also occur towards the end of the simulation.

The policy exhibits some notable qualitative behaviors. For one, it begins with a large plasma current ramp-rate, but decreases the ramp-rate as the plasma state enters the regime where the H to L mode transition is expected. This is an intuitively correct behavior for two reasons: 1) the H to L mode transition  can lead to a large $\frac{d}{dt}B_v$, and one way to decrease its value is to decrease the ramp-rate, and 2) $\beta_p$ starts approaching the user-specified limit, which happens when the current decreases faster than the stored thermal energy.

\subsection*{Feed-forward Optimization with Robustness to Physics Uncertainty}

A natural question that arises when using a trained control policy to generate feedforward trajectories is whether we might instead directly optimize feedforward trajectories, skipping the policy representation step entirely. This approach has a number of potential benefits, particularly given the risk involved in transferring an untested learned control policy to hardware.
The main difficulty that arises here is that while a feedback policy is able to observe and adjust to disturbances and uncertainty in the dynamics, purely feedforward trajectories must be designed to be robust to those disturbances \textit{without feedback}.

\begin{figure}[htbp]
    \centering
    \begin{minipage}[t]{0.36\textwidth}
        \centering
        \hspace{5mm}\textbf{PopDownGym} 
        \includegraphics[width=\textwidth]{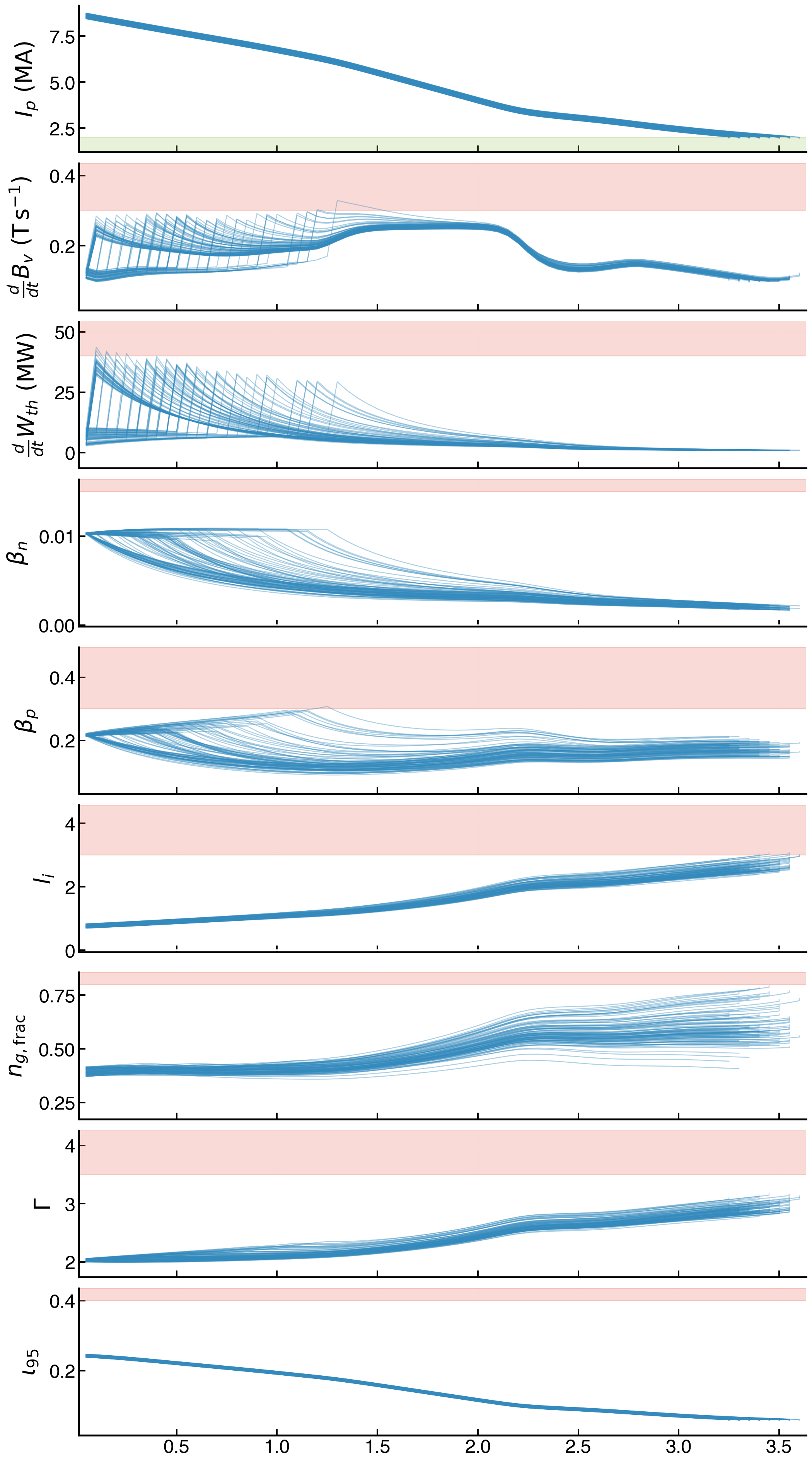}
    \end{minipage}
    \hspace{0.5cm}
    \begin{minipage}[t]{0.6\textwidth}
        \centering
        \textbf{RAPTOR} 
        \includegraphics[width=\textwidth]{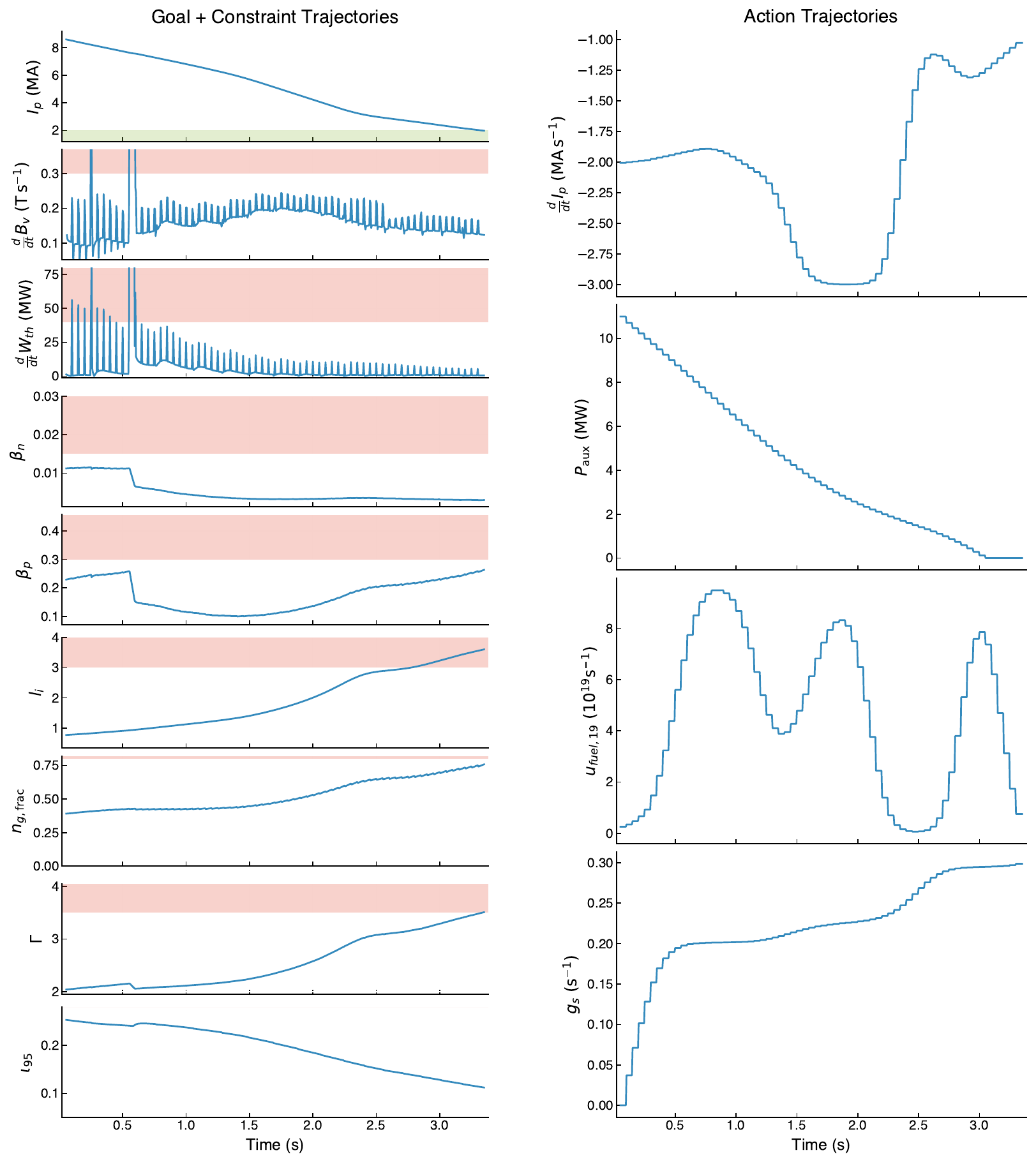}
    \end{minipage}
    \caption{(Left) results from parallel PopDownGym instances running the same feed-forward trajectory found via robust optimization. (Right) results from RAPTOR simulation running the feed-forward trajectory found via robust optimization on PopDownGym.}
    \label{fig:robust_ff}
\end{figure}

To address this challenge, we implement a robust feedforward trajectory optimization pipeline that uses a large number of parallel simulations, each with different physics assumptions (i.e. random choices for the parameters in Table~\ref{tab:random_param_ranges}). We optimize a single feedforward trajectory to minimize the average constraint violation across all of these simulations. The optimized trajectory and its performance across a range of random parameter values are shown in Figure \ref{fig:robust_ff}, showing that the optimized feedforward trajectory achieves similar hitting time and constraint satisfaction rates to the feedback policy learned using RL. The trajectory transferred to RAPTOR is also shown in Figure \ref{fig:robust_ff}, where we see it does a similarly good job of avoiding the constraint limits as the RL trained policy, with the exception of the $l_i$ limit, which it violates towards the end of the trajectory, likely due to a sim2sim gap between the training and test environments. Given that the learned policy manages to mostly avoid this limit, this results highlights the advantages of real-time feedback for making course corrections.

\subsection*{Sensitivity Analysis}
Our reinforcement learning system also enables \textit{sensitivity analysis} with respect to constraint limits and physics uncertainty. The former is enabled by the constraint-conditioned policy which allows the exploration of the optimal ramp-down for different constraint settings, and the latter is enabled by the gym environments' massive parallelization. Such information is critical for both device design and operations as it helps designers, operators, and physics modellers make decisions on how to allocate scarce resources. For example, our results in Figure \ref{fig:sensitivity_analysis} show that the optimal time to ramp-down is highly sensitive to the $l_i$ constraint, thus indicating that improvements to the vertical stability (VS) system is critical for enabling fast, safe ramp-downs. On the physics uncertainty front, Figure \ref{fig:sensitivity_analysis} indicates, for example, strong sensitivity to the H-factor, but relatively low sensitivity to the ratio of electron temperature to ion temperature. While such results require validation, they provide information on where to allocate the efforts of extremely computationally expensive simulations and device experiments.

\subsection*{Robustness Analysis}
Similar to the sensitivity analysis carried out for the constraint-conditioned policy learned via PPO, we can examine the robustness of the static feedforward trajectory to varying physical parameters. Figure~\ref{fig:feedforward_sensitivity} shows a sensitivity analysis indicating which parameters have the strongest effect on the constraint satisfaction of the feedforward trajectory; we see that H-factor has the strongest effect, with low H-factor making constraint violation more likely.

\section*{Discussion}
This paper develops a reinforcement learning approach to address the problem of disruption avoidance during tokamak ramp-downs, with a particular emphasis on the needs of operating burning plasma tokamks such as SPARC and ITER. In those contexts, it is desirable for disruption avoidance solutions to work with as few shots of training data as possible. In addition, physics models, controllers, and trajectories need to be updated as soon as possible given new data to avoid delays and enable operational continuity.

To train the policy, we propose a new sample-efficient approach to build hybrid physics and machine learning models using the frameworks of neural differential equations and differentiable simulation. The sample efficiency gained from the physics structure of the model enabled a successful demonstration of sim2sim policy transfer while  training the dynamics model on only $\approx 300$ simulations, in contrast to prior works which used order $10^4$ simulations. The model is fully implemented in JAX, enabling massive GPU parallelism across physics uncertainty. 

We also address the issue of safety-criticality for near-term burning plasma tokamaks such as SPARC and ITER and demonstrate two ways in which fusion experiments may leverage the benefits of reinforcement learning in a way more compatible with safety-critical contexts. First, we further leverage GPU parallelism during policy training to parallelize across constraint settings to train a constraint-conditioned policy. This constraint-conditioned policy enables the user to adjust constraint settings at inference time and generate new trajectories in seconds with a single policy rollout. We propose that the constraint-conditioned policy may be a useful trajectory design assistant for tokamak operators, as unexpected developments during tokamak operations often demands the adjustment of settings. Since the constraint-conditioned policy is being used to design trajectories offline, its results can be checked and validated by human operators and against validation simulations. Second, we demonstrate that the gym environment used in conjuction with other methods for feed-forward optimization. As a demonstration, we use an evolutionary algorithm to perform robust optimization of a feed-forward trajectory on parallel simulators with different physics settings. The resulting feed-forward trajectory is transferred to RAPTOR and avoids most constraint limits, but does violate the $l_i$ constraint towards the end of the trajectory by a significant amount, an issue which the policy run in closed-loop with RAPTOR largely avoids. This result highlights one benefit of having a  policy make corrections in real-time, a benefit which can then be evaluated against the risks of running a learned policy on an expensive machine.

The biggest limitation of this work is the use of a relatively low fidelity plasma simulator as a testbed, the choice of which was motivated by the rapid iteration cycles enabled by RAPTOR, which proved critical for advancing and iterating the large number of new techniques presented in this work. Ongoing work is progressing to replace the ``RAPTOR'' box used in this paper with an existing tokamak. In that setting, data from the tokamak will be used to train a dynamics model on which the control policy is then trained. Future work should also apply this approach to the highest fidelity simulations possible of upcoming devices such as SPARC and ITER. It is the hope of the authors that such a research programme will enable the practical application of the powerful tools of machine learning to make progress on the challenging and critical problem of disruption avoidance.

\section*{Methods}\label{sec:methods}
\subsection*{Reward Function}\label{subsec:reward}
The reward function is defined as:
\begin{equation} \label{eq:reward_fn}
    r(\mathbf{x}_t, \mathbf{a}_t) = \underbrace{\mathbbm{1}_{I_{p,
            t}\leq 2\text{MA}}(\mathbf{x}_t)}_{\text{Reward for reaching goal}}
    +
    \underbrace{I(\mathbf{x}_t)}_{\text{Guide term for current}} +
    \underbrace{\sum_{i=1}^{n_{cons}} b_i(\mathbf{x}_t)}_{\text{Penalty terms
        for
        constraints}}
\end{equation}
where the first term provides a fixed reward upon reaching the goal of two MA
of current, the second term $I(\mathbf{x}_t)$ rewards decreases in current to
help guide the policy towards the goal and is given by:
\begin{align}
    I(\mathbf{x}_t) = -1 + \frac{4}{\exp(0.5I_{p,t}) + 2.0 + \exp(-0.5I_{p,t})}
\end{align}
This reward term was inspired by one used for tracking error in
quadruped locomotion\cite{hwangbo2019learning} with coefficients chosen through manual tuning. Finally, the summation term consists of penalty
terms for violating user-specified constraints on undesirable limits such as
disruptive limits. In this work, all constraints in state are implemented in
the following form:
\begin{align}
    g_i(\mathbf{x}_t)\leq \mathcal{L}_i
\end{align}
where $g_i(\mathbf{x}_t)$ is the $i_{th}$ quantity to be constrained, expressed
as a function of state, and $\mathcal{L}_i$ is the user-specified limit for the
constraint. For example, the Greenwald Limit would be implemented with the
following $g$ and $\mathcal{L}_i$:
\begin{align}
    g_i(\mathbf{x}_t) = \bar{n}_e\frac{\pi a^2}{I_p}\quad \mathcal{L}_i = 1
\end{align}
There are a number of different approaches for implementing constraints in
POMDPs \cite{so2023solving,altman2021constrained}. In our formulation, each constraint is replaced with the following
``reward barrier function'', which we find to solve our problem well in practice:
\begin{align}
    b_i(\mathbf{x}_t) =
    \text{log}\left(1-\sigma\left(\text{clip}\left(\frac{g_i(\mathbf{x}_t)}{\mathcal{L}_i}\right)\right)\right)
\end{align}
where $\sigma(\cdot)$ is a sigmoid and $\text{clip}(\cdot)$ is a clip function.
This reward term is largely inspired by the logarithmic barrier functions from
classical interior point methods \cite{nesterov2018lectures} with the sigmoid
and clip functions ensuring a lower bound on reward to prevent numerical
divergence. The net effect of this function is to take on a value close to zero
when $g_i(\mathbf{x}_t)$ is away from $\mathcal{L}_i$, but then rapidly fall
off to a negative value when $g_i(\mathbf{x}_t)$ approaches $\mathcal{L}_i$.
Figure \ref{fig:reward_barrier} in the Appendix visualizes this function.
\subsection*{Dynamics Model}
\subsubsection*{Randomized Parameters}
One of the goals of this dynamics model was to explore the efficacy of training on highly simplified models, but with massively parallelized, randomized physics simulation. The randomized parameters and their domains of randomization are shown in Table \ref{tab:random_param_ranges}. In many cases, accurate simulations of these parameters would require supercomputers. Instead, we take the approach of designing policies and trajectories with robustness to variations in these parameters.
\begin{table}[h]
    \centering
    \begin{tabularx}{\textwidth}{|l|X|c|c|}
        \hline
        \textbf{Parameter} & \textbf{Description} & \textbf{Minimum Value} & \textbf{Maximum Value} \\
        \hline
        $k_{dil}$ & Main ion dilution & 0.8 & 0.9 \\
        \hline
        $k_{HL}$ & Fraction of the LH threshold at which the HL transition occurs & 0.55 & 0.75 \\
        \hline
        $H$ & H factor used in the IPB89 and IPB98 scalings & 0.8 & 1.0 \\
        \hline
        $Z_{eff}$ & Effective ion charge & 1.2 & 1.8 \\
        \hline
        $k_{te\_ti}$ & Ratio of volume-averaged electron to ion temperature & 1.0 & 1.2 \\
        \hline
        $k_{N}$ & Ratio of particle confinement time to energy confinement time & 7.0 & 9.0 \\
        \hline
        $k_{rad}$ & Ratio of total core radiated power to bremsstrahlung & 2.0 & 3.0 \\
        \hline
    \end{tabularx}
    \caption{Randomized parameter ranges used for policy training.}
    \label{tab:random_param_ranges}
\end{table}

\subsubsection*{Geometry Evolution Parameter}
To enable the full training environment to run on GPU in the absence of a GPU-capable Grad-Shafranov solver, we use a pre-computed set of ideal MHD equilibrium shapes and restrict the policy to only decide on how quickly to evolve through said shapes. To do so, we define a ``progress variable'' $g_s\in[0,1]$ where 0 corresponds to the first shape and 1 corresponds to the last. The policy then decides on the rate of change of $g_s$ at each time step. Key quantities used in the model are interpolated between the shapes and thus become functions of $g_s$. The quantities used are: $V(g_s)$, the plasma volume, $\frac{\partial V}{\partial \rho}(\rho,g_s)$, the derivative of volume with respect to the toroidal flux coordinate, $\kappa_a(g_s)$, the area elongation, $a(g_s)$, the minor radius, $\delta(g_s)$, the triangularity. From here on out, the $g_s$ argument is suppressed for notational simplicity.

\subsubsection*{Power and Particle Balance}
We make use of simple power and particle balance equations \cite{walker2020introduction}:
\begin{align}
    \frac{dW_{th}}{dt}     & = \frac{-W_{th}}{\tau_E} + P_{Ohm} + P_{\alpha} -
    k_{rad}P_{brems} + P_{aux}                                      \\
    \frac{dN_{i}}{dt} & = \frac{-N_i}{k_N\tau_E} + fuel\label{eq:particle_balance}
\end{align}
The Ohmic power, $P_{Ohm}$, alpha heating power, $P_{\alpha}$, and Bremsstrahlung radiation power $P_{brems}$ terms require profile shapes to be accurately calculated. We take the approach of mapping volume averaged quantities to profiles shapes learned via principal components analysis (PCA) on the RAPTOR dataset. First, volume averaged electron and ion temperatures and densities are obtained from state variables as such:
\begin{align}
    \frac{2W_{th}}{3V} &= \langle p\rangle  \approx \langle T_e\rangle \langle n_e\rangle + \langle
    T_i\rangle\langle n_i\rangle =\langle T_e\rangle\frac{\langle
        n_i\rangle}{k_{dilution}} + \frac{\langle T_e\rangle}{k_{te\_ti}}\langle
    n_i\rangle\\
    \langle n_i\rangle &= \frac{N_i}{V}, \quad \langle T_e\rangle = \frac{2W_{th}}{3V}\left(\frac{\langle n_i\rangle}{k_{dilution}} + \frac{\langle n_i\rangle}{k_{te\_ti}}\right)^{-1}, \quad \langle n_e\rangle = \frac{\langle n_i\rangle}{k_{dilution}}, \quad \langle T_i\rangle = \frac{\langle T_e\rangle}{k_{te\_ti}}
\end{align}
Observe that the only unknown in the set of equations above is the volume-averaged electron temperature $\langle T_e\rangle$. We represent each of the four profiles with two single-component PCA basis: one for H-mode and one for L-mode. Below, the $T_e$ profle is used as an example, but the same methodology applies for the other profiles:
\begin{align}
    T_e(\rho) = cv(\rho) + b(\rho)
\end{align}
where $v(\rho)$ and $b(\rho)$ are learned via PCA on the RAPTOR simulation dataset. The volume integral and its approximation with Legendre-Gauss quadrature can then be computed as:
\begin{align}
    \langle T_e\rangle = \frac{1}{V}\int_0^1 \left[cv(\rho) + b(\rho)\right]\frac{\partial V}{\partial \rho}(\rho) d\rho\approx \frac{1}{\sum_{j=1}^{n_{gauss}} w_j\pder[V]{\rho}(\rho_j)}\left[c\sum_{j=1}^{n_{gauss}}w_jv(\rho_j)\pder[V]{\rho}(\rho_j) + \sum_{j=1}^{n_{gauss}}w_jb(\rho_j)\pder[V]{\rho}(\rho_j)\right]
\end{align}
where $n_{gauss}$ is the number of Legendre-Gauss points, $w_j$ are the weights, and $\rho_j$ are the Legendre-Gauss points. Rearranging the above equation for $c$ we get:
\begin{align}
    c \approx \frac{\langle T_e\rangle \sum_{j=1}^{n_{gauss}}w_j\pder[V]{\rho}(\rho_j) - \sum_{j=1}^{n_{gauss}}w_jb(\rho_j)\pder[V]{\rho}(\rho_j)}{\sum_{j=1}^{n_{gauss}}w_jv(\rho_j)\pder[V]{\rho}(\rho_j)}
\end{align}
Now everything on the right hand-side is a function of state or a pre-computed quantity. Since $v(\rho)$ and $b(\rho)$ are constant when running the simulation, $c$ fully parameterizes the $T_e(\rho)$ profile.
\subsubsection*{H-mode and L-mode}
We use the IPB98 scaling for H-mode \cite{ITER_1999} and IPB89 \cite{kaye1997iter} for L-mode. The H to L mode transition threshold is subject to considerable uncertainty, making our approach of randomizing across physics uncertainty particularly relevant. In this model, we treat the HL back-transition threshold as a constant multiple of the established the LH threshold from \cite{martin2008power}, with the constant randomized across simulations:
\begin{equation}
    P_{HL} = k_{HL} 2.15e^{0.107} n_{e,20}^{0.782} B_T^{0.772} a^{0.975} R^{0.999}
\end{equation}

\subsubsection*{Internal Inductance Evolution}
Authors in\cite{romero2021model} derived a simple three-state ODE system for the coupled internal inductance and plasma current evolution. We make use of their equations for the internal inductance and plasma current evolution, which are exact, and replace their voltage dynamics, which are ad-hoc, with a neural network, resulting in the following ODE system:
\begin{align}
    \frac{dL_i}{dt} & = \frac{2}{I_p}(-V_{ind} - V_R)                  \\
    \frac{dI_p}{dt} & = \frac{1}{L_i}(2V_{ind} + V_R)                  \\
    \frac{dV_R}{dt}   & = NN(I_p, L_i, V_R, V_{ind}, \langle T_e\rangle)
\end{align}

\subsection*{Algorithms}
\subsubsection*{Optimizing Constraint-Conditioned Policies with Proximal Policy Optimization}
We apply PPO \cite{schulman2017proximal}, a state of the art on-policy reinforcement learning algorithm to solve for the constraint-conditioned policy that maximizes the reward function \eqref{eq:reward_fn}. A fully connected network with three hidden layers of $256$ units each is used for both the policy and value networks with hyperbolic tangent activation functions. The environment observations are concatenated with the desired constraint values to form the input vector for the policy network. The value network additionally takes the current simulation parameters as input to improve performance \cite{pinto2017asymmetric,zhu2018reinforcement}. All input observations and output actions are normalized to the range $[-1.0, +1.0]$.

We exploit the capabilities of JAX and collect transitions from $2048$ environments in parallel. During each training step, it took $0.652$ ($\pm 0.024$) seconds to collect experience and $0.048$ ($\pm 0.002$) seconds to update the policy and value networks. Training converged in under two hours on a single Nvidia RTX 4090 GPU.

\subsubsection*{Robust Feedforward Trajectory Optimization with Evolutionary Strategies}

We apply a highly parallelizable black-box optimization algorithm known as evolutionary strategies (ES) to optimize robust feedforward trajectories~\cite{salimans2017evolution}. In contrast to policy gradient methods like PPO, which consider the expected discounted reward in Eq.~\eqref{eq:pomdp_objective}, ES requires an end-to-end objective function that assigns a reward to a given fixed-horizon trajectory; we maximize:
\begin{equation}\label{eq:es_objective_dense}
    U_{dense}(\mathbf{a}_{0:T}) = \mathbb{E}_{\theta} \left[ \frac{1}{T} \sum_{t=0}^T r(\mathbf{x}_t, \mathbf{a}) \right]
\end{equation}
where the expectation is with respect to the random parameters $\theta$ sampled uniformly over the ranges given in Table~\ref{tab:random_param_ranges}.

The feedforward trajectory is represented as a cubic interpolation between 10 equally spaced control points for each action. Trajectories are rolled out for 100 timesteps with 0.05 s timestep. We use a population size of 256, approximate the expectation in Eq.~\ref{eq:es_objective_dense} using $10^3$ random samples, and run the evolutionary strategy for 250 generations. The learning rate is initialized at $10^{-1}$ and decays with rate $0.995$ to a minimum of $10^{-3}$. We use the ES defined by Salimans \textit{et al.}\cite{salimans2017evolution} as implemented in JAX in the \texttt{evosax} library\cite{evosax2022github}.

\subsection*{RAPTOR Simulation Settings}
RAPTOR is configured to evolve the $T_e, T_i$, and current profiles by solving the corresponding transport equations. Following prior work, the particle transport equations are not solved \cite{teplukhina2017simulation}, but rather a separate instance of the simple particle balance model \ref{eq:particle_balance} is implemented RAPTOR side and receives the fueling commands. This separate model is configured with the maximum $k_N$ value of 9.0 used in the training model to challenge the policy as gas injection can be used to directly slow down the density decrease, but actuators that can directly speed up the density decrease do not exist. This approach is motivated by the challenges of modelling plasma density dynamics\cite{walker2020introduction}; the goal instead is to provide reasonable density set-points for real-time density controllers. The H-L back-transition is triggered when the RAPTOR simulated power conducted to scrape-off layer falls below the HL back-transition threshold. 

\section*{Data and Code Availability}
The gym environment, trained policies, trajectories, and scripts to generate figures will be available at the time of publication at \url{github.com/MIT-PSFC/PopDownGym}. Code used for RAPTOR simulations will not be available due to the closed-source status of RAPTOR at the time of this work.

\bibliography{references}

\section*{Acknowledgements}
This work was funded in part by Commonwealth Fusion Systems. The authors would like to thank Federico Felici and Simon Van Mulders for help with RAPTOR and technical discussions. The authors would also like to thank colleagues at the MIT Plasma Science and Fusion Center (PSFC) and Commonwealth Fusion Systems (CFS) for helpful technical discussions.

\section*{Author contributions statement}
Allen M. Wang led the project, developed the Gym environment, the initial PPO implementation, RAPTOR transfer, and wrote most of the paper. Oswin So developed the high performance final PPO implementation, trained the constraint conditioned policies, and led visualizations. Charles Dawson developed the robust optimization algorithm and benchmarked the Gym environment. Darren Garnier conceptualized the project and helped Allen M. Wang get up to speed on plasma physics and fusion. Cristina Rea advised the project on the fusion and disruptions front, and revised the paper. Chuchu Fan advised the project on the controls and reinforcement learning front.
\section*{Additional information}
\subsection*{Competing Interests}
Allen M. Wang, Darren Garnier, and Cristina Rea recieved funding from Commonwealth Fusion Systems (CFS) and have worked directly with CFS on the SPARC project.
\appendix
\newpage
\section{Sensitivity Analysis}
\begin{figure}[!h]
    \centering
    \includegraphics[width=\linewidth]{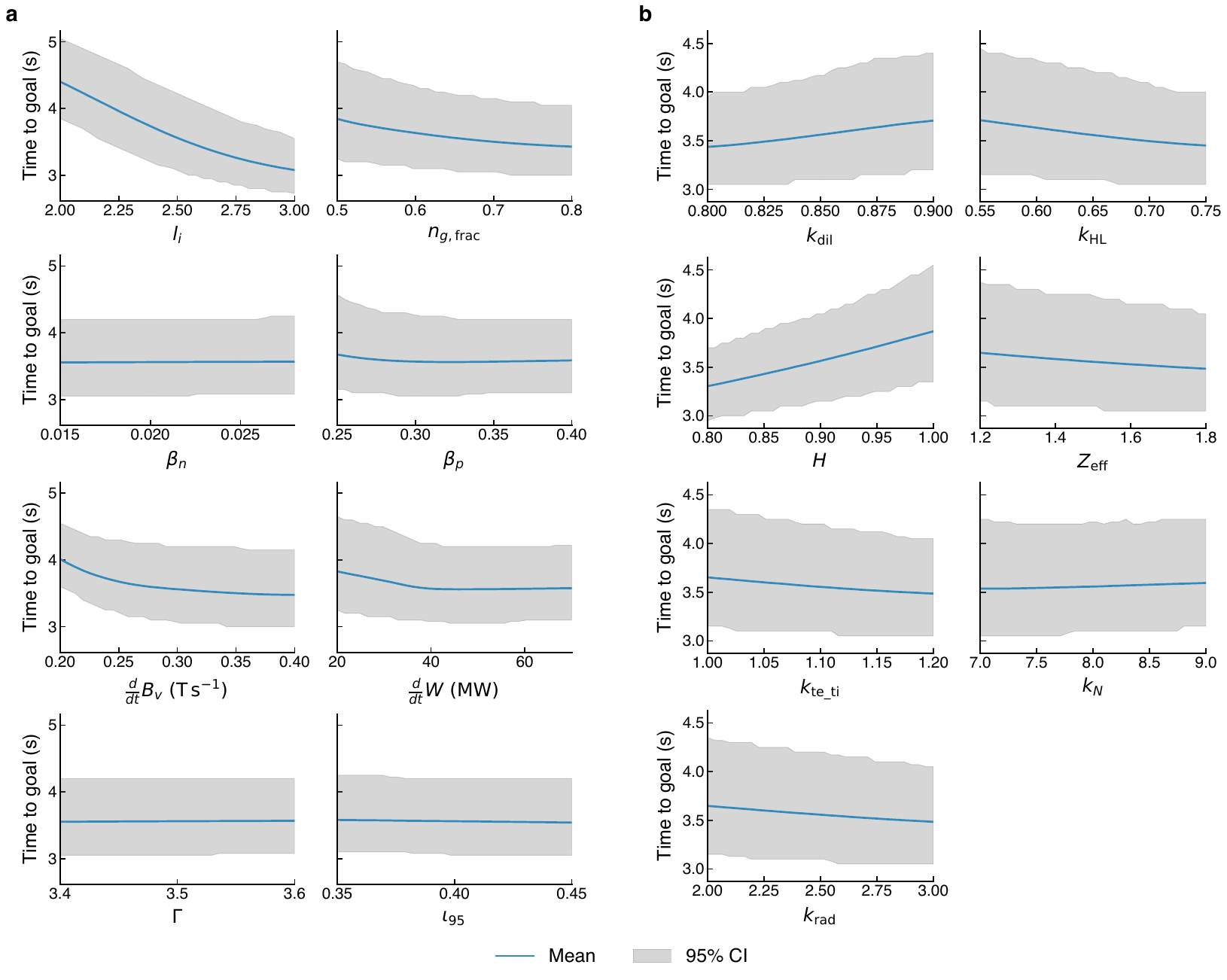}
    \caption{\textbf{a,b}, Sensitivity of the time to goal with respect to constraint limit settings (\textbf{a}) and physics uncertainty (\textbf{b}), with the $95\%$ confidence intervals corresponding to the physics uncertainty. In \textbf{b}, the given physics parameter is fixed while the other physics uncertainty parameters are sampled randomly.
    }
    \label{fig:sensitivity_analysis}
\end{figure}

\section{Robustness Analysis}
\begin{figure}[!h]
    \centering
    \includegraphics[width=\linewidth]{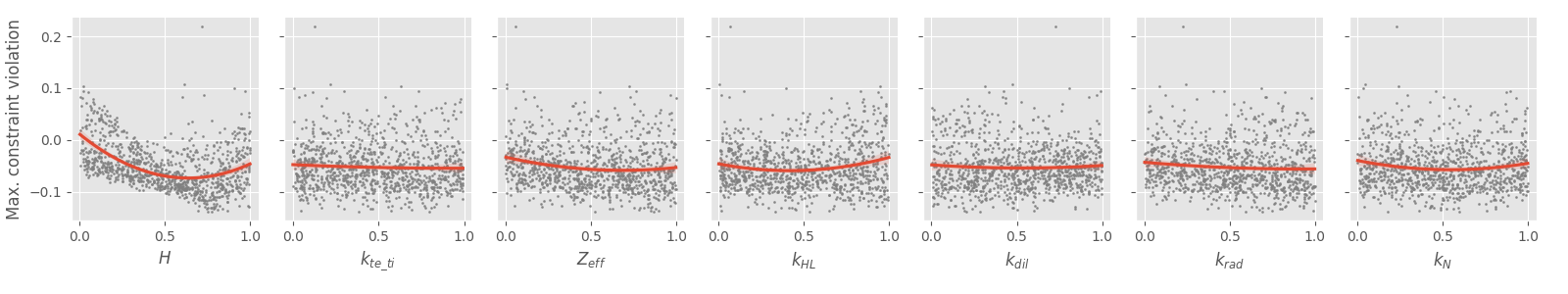}
    \caption{Sensitivity analysis showing the effect of varying physical parameters on constraint satisfaction for the feedforward trajectory. Raw data points are shown in grey, and a quadratic best fit and 95\% confidence interval are shown in orange.}
    \label{fig:feedforward_sensitivity}
\end{figure}

\section{Dynamics Model Performance}
\begin{figure}[!h]
    \centering
    \includegraphics[width=0.5\linewidth]{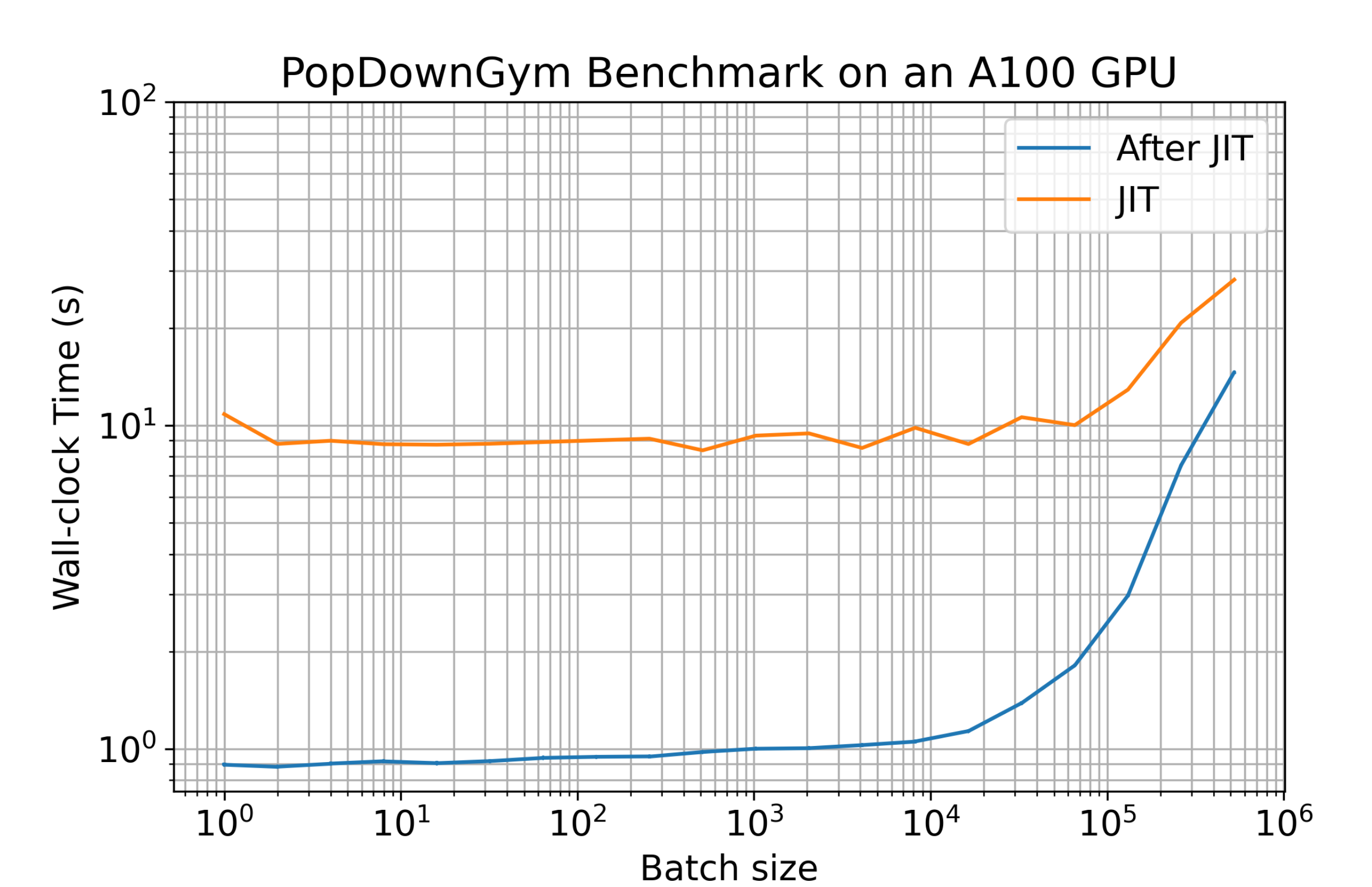}
    \caption{Wall-clock time as a function of number of simulations per
        vectorized batch showing how GPU parallelism enables $10^4$ simulations
        in
        approximately a second, where each simulation consists of 100 gym
        environment
        time steps. The benchmark is performed with 101 trials for each batch
        size; the
        first trial includes just-in-time (JIT) compilation time and is shown
        in
        yellow. The average of 100 trials after JIT is shown in blue.}
    \label{fig:gym_benchmark}
\end{figure}

The entire dynamics model is implemented in Jax. Thus, in addition to being
fully differentiable, the model also runs on GPU, enabling massive
parallelization. Figure \ref{fig:gym_benchmark} shows benchmark results of
simulation

\section{Visualizing the Reward Barrier Function}
\begin{figure}[!h]
    \centering
    \includegraphics[width=0.75\textwidth]{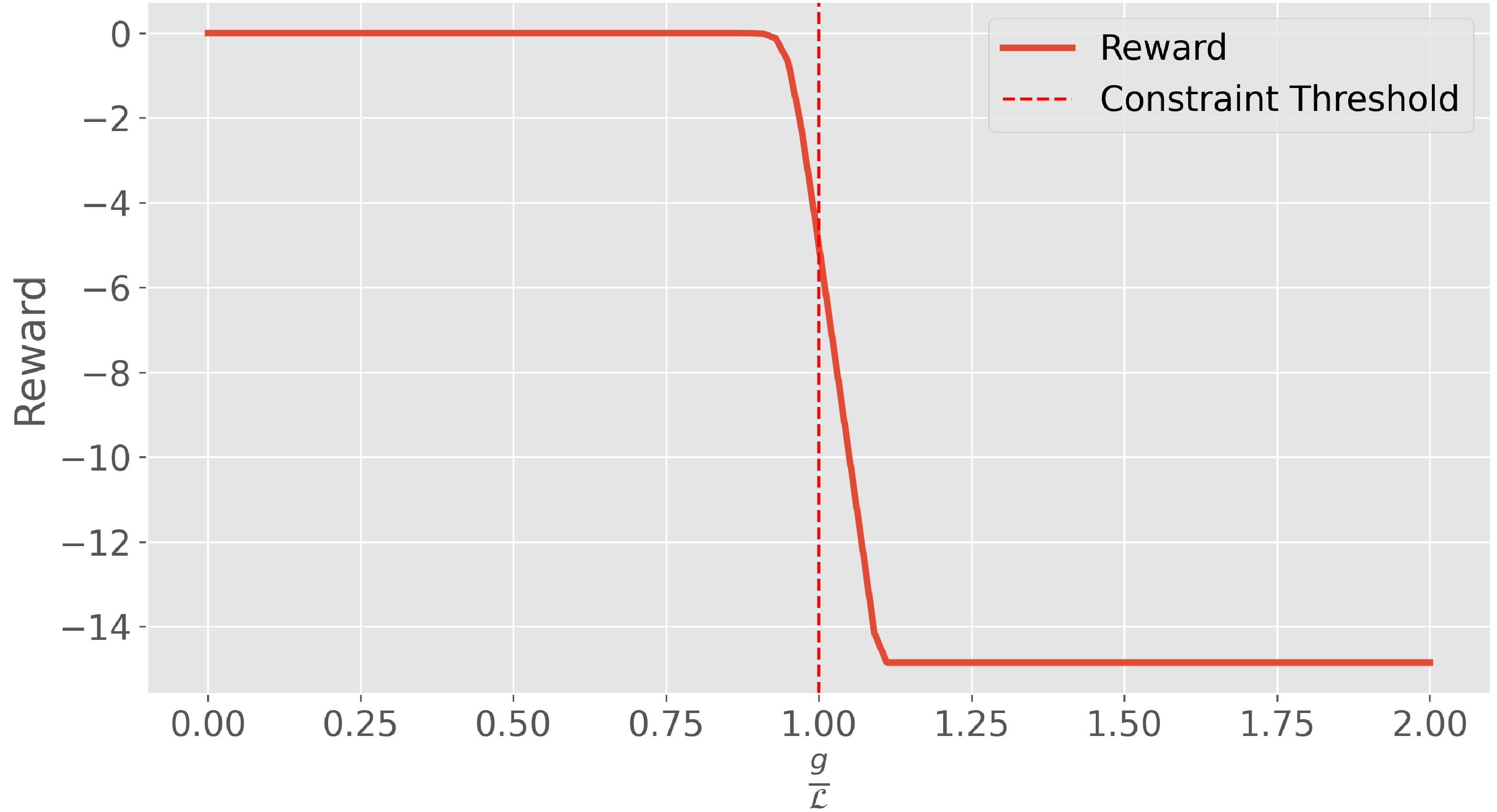}
    \caption{Visualization of the reward barrier function}
    \label{fig:reward_barrier}
\end{figure}

\end{document}